\documentstyle[12pt,aasms4]{article}
\def \kms{km~$\rm{s}^{-1}$}
\def \cc{$\rm{cm}^{-3}$}
\def \lam{$\lambda$}

\def \mum{$\mu$m}

\newfont{\rten}{cmr10} 

\def\arcmin{\hbox{$^\prime$}}
\def\arcsec{\hbox{$^{\prime\prime}$}}

\begin{document}

\title{Optical and Near Infrared Study of the Cepheus E outflow, a
very low excitation object.}
  
\author{S. Ayala}
\affil{Instituto de Astronom\'\i a,
UNAM, Ap. 70-264, 04510 M\'exico, D. F., M\'exico}
\affil{sayala@astroscu.unam.mx}
  
\author{A. Noriega-Crespo}
\affil{SIRTF Science Center, IPAC, Caltech 100-22, Pasadena, CA 91125, USA}

\author{P. M. Garnavich}
\affil{Harvard-Smithsonian Center for Astrophysics, 60 Garden St. (MS 42),
 Cambridge, MA 02138}

\author{ S. Curiel and A.C. Raga}
\affil{Instituto de Astronom\'\i a,
UNAM, Ap. 70-264, 04510 M\'exico, D. F., M\'exico}

\author{K.-H. B\"ohm}
\affil{Astronomy Department, University of Washington, Seattle,
98195}

\and

\author{J. Raymond}
\affil{Harvard-Smithsonian Center for Astrophysics, 60 Garden St. (MS 42), 
Cambridge, MA 02138}

\begin{abstract}
We present images and spectra of the Cepheus E (Cep E) region
at both optical and infrared wavelengths.
Only the brightest region of the southern lobe of the Cep E 
outflow reveals optical emission, suggesting that the extinction close
to the outflow source plays an important r\^ole in the observed
difference between the optical and IR morphologies.
Cep E is a unique object since it provides a link between the spectroscopic
properties of the optical Herbig-Haro (HH) objects and those of deeply
embedded outflows.

The observed H$_2$ infrared lines allow us to determine an excitation
temperature of $\sim 2300$~K, an Ortho-to-Para ratio of $\sim 3$,
and an H$_2$ (1,0)/(2,1) S(1) line ratio of $\sim 9$. These results
are consistent with the values observed for HH objects with detected
NIR emission lines, with shock excitation as the main mechanism for 
their formation, and also with the values observed for embedded,
NIR flows.

The optical spectroscopic characteristics of Cep E (HH~377) appear to be 
similar to the ones of low excitation HH objects.
However, the electron density determined from the [SII]6731/6717 line
ratio for this object ($n_e=$ 4100 cm$^{-3})$, and the [OI]6300/H$\alpha$,
[SII](6717+6731)/H$\alpha$ ratios  are higher than the values of all 
of the previously studied low excitation 
HH objects. This result is likely to be the consequence of an anomalously 
high environmental density in the HH~377 outflow.

The ionization fraction obtained for HH~377 is $x_e \sim 1\%$
From this result, together  with the observed [OI]6300/H$\alpha$ line
ratio, we conclude that the observed H$\alpha$ line emission is 
collisionally  excited.
From a comparison with shock models, we also conclude that the extinction 
towards  HH~377 is very low.
Comparing the observed H$\beta$ and H$\alpha$ fluxes of HH~377 with model
predictions,
we determine a shock speed  between 15 and 20 km s$^{-1}$, although 
somewhat higher velocities also produce spectra with line ratios 
that qualitatively agree with the observations of HH~377.
\end{abstract}

\keywords{jets and outflows -- infrared, optical: interstellar: lines --
ISM: individual objects: Cepheus E, HH~377}

\section{Introduction}

The Cepheus E (Cep E) outflow was first detected in the $^{12}$CO J=1-0 
transition in some of the early radio studies of star formation in molecular 
clouds \markcite{sar77}. 
In his catalog of molecular outflows, Fukui (1989) first pointed out the 
presence of a bipolar, high velocity outflow in this region.
The bipolar nature of Cep E became clear with the K$^\prime$ image 
of the molecular outflow obtained by Hodapp (1994), which included the
stronger NIR H$_2$ lines, and revealed a relatively compact system
(with a size of $\sim$ 1.5\arcmin).
Subsequent studies in the near/mid infrared have shown that the outflow is
quite bright in the H$_2$ (1,0) S(1) 2.12 $\mu$m~line, consistent with 
models of shock excited H$_2$ gas \markcite{eis96,lad97,nor98}. 
Suttner et al. (1997) have used three-dimensional hydrodynamic simulations 
of highly collimated  molecular outflows in order to model the morphology 
of Cep E. These simulations, however,
required a very high density {\it in} the jet ($10^5$ cm$^{-3}$), which are 
not consistent with the hot and dense CO bullets recently found in the flow
\markcite{hat99}, with densities of $\sim 10^4$ cm$^{-3}$.

The IRAS 23011+6126 source was originally identified as the main candidate
for the outflow source. However, the presence of multiple outflows in near 
infrared and radio wavelengths \markcite{eis96,lad97} indicates the existence 
of at least two sources, which have recently been confirmed by OVRO 
observations at 1.3 and 2.6mm \markcite{tes98}.
The sources are embedded and invisible at optical and near infrared
wavelengths, and are likely to be Class I or Class 0 protostellar 
objects \markcite{lef96,and93}.
In addition, Noriega-Crespo et al. (1998)  detect one source
at 6.9 $\mu$m using ISOCAM, which is well detected in all IRAS bands.

The present study has been motivated by the detection of emission
at optical wavelengths in a small section of the southern lobe of the 
Cepheus E outflow by Noriega-Crespo (1997) and Devine et al. (1997).
Noriega-Crespo (1997) mentions that H$\alpha$ and [SII] 6717/31 images 
reveal a compact knot which is the optical counterpart of the southern 
bowshock observed at 2 $\mu$m by Eisl\"offel et al. (1996). This optical
knot has been named HH~377 \markcite{dev97}.

In this study, we explore the link between the physical 
properties of the outflow as determined from optical imaging and 
spectroscopy, and compare these results with those obtained from
observations in the near infrared. Our goal is to understand the
development of very young stellar outflows (we notice that Cep E has
a dynamical age of $\sim 3\times 10^3$ years (Noriega-Crespo et al. 1998))
and the relationship between the mechanisms that produce the
infrared and optical emission.

The paper is organized as follows.
In Section 2 we describe the different observations obtained for this work, 
and comment on the reduction and calibration techniques. In Section 3,
we present the results obtained from our infrared and optical observations.
Finally, in Section 4 we compare the physical properties of Cepheus E
deduced from the optical and  the NIR observations with other Herbig Haro
objects.

\section{Observations}
The optical and infrared observations were carried out at three different
observatories. The log of the observations is presented in 
Table 1, and they are described  in detail below.

\subsection{Near Infrared Imaging and Spectroscopy}

The NIR images of Cep E were obtained at two observatories. A set of images
was obtained at the Apache Point Observatory 3.5m telescope (APO 3.5m) with a
$256\times 256$ array at f/5 and a 0.482\arcsec~per pixel scale. 
The central wavelengths (and bandpass) of these images were at 2.12\mum~
(1$\%$), for the H$_2$ (1,0) S(1)  line, and 2.22\mum (4$\%$) for the 
nearby continuum.

Another set of images was obtained at the Observatorio Astron\'omico Nacional
at San Pedro M\'artir 2.1m telescope (OAN SPM 2.1m) with the IR
camera/spectrometer CAMILA \markcite{cru94}, which has a $256\times256$ array 
providing a 0.85 \arcsec/pixel scale at f/4.5. The central wavelengths 
(and bandpass) of the filters used to obtain these images were 
2.12\mum~(1$\%$), 2.25\mum~(1$\%$) for the H$_2$ (2,1) line, cK (2$\%$) and 
Br$\gamma$ 2.16\mum.

The frames were flattened with  a combination of low and high
illumination sky flats obtained at sunset. 
The data were processed by subtracting a median-filtered image of
nearby sky frames taken with the same integration time and with 
offsets between 30 and 100\arcsec. Bad pixels were removed with standard
techniques.

The processing of our frames was done with Image Reduction and Analysis 
Facility (IRAF)\footnote{IRAF is distributed by the National Optical 
Astronomy Observatory (NOAO), which is operated by the Association of 
Universities for Research in Astronomy (AURA), Inc. under cooperative 
agreement with the National Science Foundation} based 
programs. For each filter, 10 (APO data) and 9 (OAN SPM data) overlapping
frames were taken (see Table~1 for the integration times).
The frames were aligned using several field stars, and median combined into 
the final images. The image scale and orientation were calculated using
7 common stars between the IR frames and an optical image from 
the Digitized Sky Survey, with a resultant mean uncertainty of less than 
one arcsecond in the positions.

The NIR spectra  were obtained with the Multiple Mirror Telescope Observatory 
(MMTO). The set of H and K band  spectra of the North and South lobes of 
Cep E were taken with the Rieke FSPEC IR spectrometer.
The slit was aligned in the E-W direction in both lobes (see Figure 1)
and alternate exposures were chopped between the source and the local sky.  
The atmospheric absorption and sensitivity variation
along the dispersion were corrected by observing a bright late-F type
star at nearly the same airmass as the target data. 
The spectra were obtained with a 1.0$\arcsec$ wide slit, and the 
extraction was done over an area of 4.0 $('')^2$ for each outflow lobe.

\subsection{Optical Imaging and Spectroscopy}

As mentioned above, the south lobe of Cep E is detected at optical 
wavelengths,  in the H$\alpha$  and [S~II] 6717/31 lines \markcite{dev97}.
Our narrow band images in H$\alpha$ and [S~II] were obtained with the
1.2m telescope at Fred Lawrence Whipple Observatory (FLWO 1.2m). 
A thinned, back-side illuminated, AR coated Loral $2048 \times 2048$ CCD was 
used with a plate scale of 0.315 \arcsec/pix. The central wavelength and
FWHM of the narrow band H$\alpha$, [S~II] and continuum filters are,
respectively, $\lambda$6563, 25 \AA, $\lambda$6724, 50 \AA, and
$\lambda$6950, 400 \AA.
The CCD images were binned two by two giving a 0.63~\arcsec/pix scale, and were
processed using IRAF, in the standard way. For each filter, the final 
image corresponds to the median-filter of three 600 seconds frames.

A low resolution, long-slit spectrum over a $\sim 4500-7000$ \AA~ 
wavelength range with $\simeq 2$ \AA/pix was obtained at the MMTO 
telescope. The spectrum was reduced using IRAF and flux calibrated using 
the standard star HR8687.

Two  long-slit spectra were also obtained at the FLWO 1.5m telescope with 
the FAST spectrometer, using a Loral $512 \times 2688$ coated CCD with 
15$\mu$m pixels.
Two different spectral resolutions (1.49 \AA/pix and 0.75 \AA/pix) 
 were used to cover the wavelength
ranges of  5500-7000 \AA\ and 6200-6800 \AA. The slit width was
1.1\arcsec~and the slit was oriented N-S. The spectra were flux calibrated
with the standard stars  BD284211 and G191B2B.

\section{Results}

\subsection{Infrared excitation}

The complex structure shown by Cepheus E in   vibrationally excited 
molecular hydrogen lines has been described in previous papers 
\markcite{hod94,eis96,lad97}. Figure~\ref{fig_ir} shows a 
section of our H$_2$ (1,0) S(1) continuum subtracted image, in which we 
plot the slit positions used to obtain the infrared spectra.

Using the H$_2~ \lambda$2.121$\mu$m and $\lambda$2.248$\mu$m images we 
construct a (1,0)/(2,1) S(1) line ratio image. The line ratio is nearly 
constant throughout the outflow with a mean  value of $10 \pm 5$. This 
result is completely consistent with previous results on this object 
\markcite{eis96}.

In our Bracket $\gamma$ + continuum image we have detected emission extending
over both lobes. The distribution of this emission is very similar to the
one of the continuum cK frame. 
Because of this similarity we suspect that the flux detected in our 
Br$\gamma$ frame is on the whole continuum emission. This interpretation is 
consistent with the absence of Br$\gamma$ line in the spectra obtained for 
the brightest knots of both lobes (see Figure \ref{fig_irspec}).

The spatially integrated spectra in the H and K bands of the northern and 
southern lobes of Cepheus E are shown in Figure \ref{fig_irspec}.
These spectra were constructed by integrating along the slit over 
the width of the lobes ($\sim 4$\arcsec) for the slit positions indicated 
in Figure \ref{fig_ir}. The identified and measured H$_2$ lines in these 
bands are indicated on the plots at the expected wavelengths.
Note that no Br$\gamma$  ($\lambda 2.166 \mu$m) line emission is detected 
in these slit positions, which correspond to the brightest knots of each lobe.
Table~2 lists the H$_{2}$ transitions (column 1), their wavelengths (column 
2; see Black \& van Dishoeck, 1987); the energy $E(v,J)$ of 
the upper level (column 3, Dabrowski 1984), and the measured 
fluxes of the identified H$_{2}$ lines (columns 4 and 5 for the N and S lobes,
respectively). From these spectra we determine a $\Delta V_r \sim$ 90$\pm 30$ 
km s$^{-1}$ between the northern and southern lobes, and the south lobe shows 
a blueshift.
Comparing the fluxes for (1,0) S(1) and (2,1) S(1) lines from our spectra
with previous values obtained from surface photometry  for these transitions
\markcite{eis96}, we find that our (1,0) S(1) flux values are lower with respect 
of the Eisl\"offel et al. values, by a factor $\sim 2$. Whereas in the case 
of (2,1) S(1) fluxes the present and previous measurements are completely 
consistent.

We have estimated the column densities $N(v,J)$ assuming that the lines are 
optically thin, in order to determine the excitation temperature, $ T_{exc}$,
of the warm molecular gas in each lobe. Under conditions of
local thermodynamic equilibrium (LTE), the relationship between
column density and energy of the upper level $E(v,J)$, is  
$ln [N(v,J)/g_{J}] = E(v,J)/k T_{exc} + C $;
where $g_{J}$ is the degeneracy of the corresponding level, $k$ is the 
Boltzmann constant, and $C$ is a constant.

Figure \ref{fig_texc} shows the excitation diagrams for Cep E,
 South and North lobes. 
We plot  $N(v,J)/g_{J}$ versus energy $E(v,J)$ of the level
for each lobe using the  fluxes listed in Table~2\footnote{The 
excitation diagrams do not include the very weak 
H$_2$ (3,2) S(5) line
presented in Table 2. This transition shows a bigger column density than 
expected, which  causes great dispersion in the estimation of
the excitation temperature, so we have excluded it from our analysis.}. 
In each panel, the transitions between the (1,0) vibrational levels 
are plotted 
with solid squares, the ones between the (2,1) levels with triangles and the 
(3,2) levels with solid circles.
The excitation temperatures, T$_{exc}$ (calculated from linear fits), are
$2340 \pm 100$ for the South lobe and $2260 \pm 110$ for the
North lobe.

In the case of a purely thermal population caused by a shock, it is
expected that a single smooth curve should fit all of the transitions in 
the excitation diagrams \markcite{gre94,sch95}, meaning that a single temperature
describes the excitation. On the other hand, in the case of 
fluorescent  H$_2$ emission, the points within vibrational levels are
expected to be aligned on separate curves \markcite{bur90}, resulting in 
different temperatures for different vibrational levels. Intermediate cases 
in which there is a combination of both excitation mechanisms are also 
possible (Fernandes \& Brand 1995; Fernandes, Brand \& Burton 1995).

The derived excitation temperatures, T$_{exc}$, for both lobes of Cepheus E 
are consistent with the  T$_{exc}$ measured  in HH objects (T$_{exc} \sim 
2000$ K, Gredel 1994), whereas  the excitation temperatures 
observed in  collisionally plus fluorescent excited objects have T$_{exc} 
\sim  3000$ values (Fernandes, Brand \& Burton 1997). However, in Figure \ref{fig_texc}
there are real dispersions around the dashed lines drawn in 
each panel. If we fit the points from different vibrational levels separately,
there are slight differences between these fits and the fit obtained
considering all of the vibrational levels together.

We have computed the rotational excitation temperatures using 
the transitions within different vibrational levels. The resulting temperatures
are listed in Table~3. In order to explore the excitation mechanisms
occurring in the Cep E outflow, we have computed some interesting
line ratios (with the fluxes presented in Table~2), which are also listed in
Table~3. This table gives the vibrational levels (column 1),
the rotational levels of the transitions used for the ratios (column 2), the
rotational temperature T$_{rot}$ derived from these transitions (column 3),
the empirical Ortho/Para ratio derived for each vibrational level (column 4),
and the (1,0)/(2,1) S(1) ratio derived for each spectrum (column 5).

As we can see, for both lobes of Cepheus E the rotational excitation 
temperatures (Table 3) are lower than the calculated vibrational
temperature T$_{exc}$ (see Figure \ref{fig_texc}). Though different
from T$_{exc}$, the rotational temperatures are
still consistent with the temperatures observed in collisionally
excited objects. We should note that the vibrational level (2,1)
shows the lowest rotational temperature.

We find that the values obtained for the 
Ortho/Para ratio are similar for the two lobes of Cepheus E.
For the Northern lobe there is a higher dispersion
for this ratio between the different vibrational levels.  
This is a result of the dispersion in the points
corresponding to the (2,1) and (3,2) levels (see top panel of Figure 
\ref{fig_texc}). This dispersion is probably a result of the fact
that the spectrum of the Northern lobe is fainter.
We conclude that the  Ortho/Para ratios estimated for Cep E
are completely consistent with 
the LTE value of 3, as expected for collisional excitation.

We analyzed the behavior of the
(1,0)/(2,1) S(1) line ratio. Column (5) of Table 3
gives the $8.50 \pm 1.10$ and $9.40 \pm 1.30$
 values obtained for the Northern and Southern lobes,
respectively. These  values are consistent with 
the result obtained from the (1,0)/(2,1) S(1) ratio image (see above),
and slightly lower than the ones measured from (1,0)/(2,1) S(1) frames by
Eisl\"offel et al. 1996.
Also, the obtained (1,0)/(2,1) S(1) line ratio is completely
consistent with the value of 10 expected for collisional excited
H$_2$ lines, and does not approach the $\sim$ 1.7
value expected for fluorescent excitation \markcite{bur90}.

A detailed analysis on the K-band spectra of Cep E has already
been performed by Ladd \& Hoddap (1997). They found that a C-type shock
with 35 km/s can explain the K-band H$_2$ fluxes and ratios.
This is consistent with the ISO Long Wavelength Spectrometer
observations, which show several emission lines from H$_2$O transitions
\markcite{nor00b}, as it is expected from molecular C-type shocks \markcite{ka96a}.

In Cep E we find a similar situation as in other  Herbig-Haro
objects (e.g. HH 1-2; Davis et al. 1994; Noriega-Crespo \& Garnavich 1994)
where J-type (inferred from the optical spectra) and C-type (inferred from the
IR spectral) seem to coexist spatially. The case of Cep E is particularly 
interesting since it appears to be a very young outflow, based on its compact 
size and the age of its exciting source. If so, a better interpretation of 
its spectral properties may require molecular time dependent shock models, 
which are currently under development \markcite{flo99}.

\subsection{The optical morphology}

In contrast with the complex structure of the H$_{2}$ outflow, the optical 
emission consists of only one compact knot (HH~377), which approximately
coincides with the brighter H$_2$ knot of the southern lobe  
\markcite{nor97,dev97}. In Figure 4 we present an H$_2$ 1-0 S(1)
contour map of Cepheus E, overlaid on grey-scale and contour
representations of our optical images.
Panel $a)$ shows grey-scale representations of
the H$\alpha$ image and panel $b)$ of the [S~II]6717,6731 image.
Panels $c)$ and $d)$ show the H$\alpha$ and [S~II]6717,6731 images
(respectively) as contour maps, over a smaller angular area.
These maps clearly show the coincidence between the optical knot and 
the brightest region of the southern lobe of the IR flow.

In Figure 4, we see that a star still remains in the H$\alpha$ continuum
subtracted image (panel $a$), to the west of HH~377.
This stellar emission could indicate either that the star has
intrinsic H$\alpha$ emission, or that the stellar continuum has
a substantially different slope than the continua of the stars
used to obtain the relative scaling between the H$\alpha$ and
the continuum frames.

The H$\alpha$  and [S~II] morphologies of HH~377 are different, the knot is 
resolved in both images, but it is clearly brighter and more compact
in the [S~II] image. In our [S~II] image, HH~377 has an angular size of
5\arcsec, resulting in a diameter of 0.02 pc (at a distance of 0.75 kpc,
Hodapp 1994) for the knot.
We can also see  that the morphology of the southern lobe of Cep E 
 is qualitatively consistent with a bow-shock model, 
in which the wings have a stronger contribution of H$_2$ emission, while
the head is dominated by the atomic/ionic emission.

Finally, in Figure 4 (panel $b$), we show the two slit positions
used to obtain the spectra described in the following section.

\subsection{Optical spectroscopic characteristics}

We have obtained two spectra (intermediate and low resolution) with the
N-S slit (see Figure 4), and a single (low resolution) spectrum with
the E-W slit (see the discussion in section 2.2). Our spectrophotometric 
data gives us the optical excitation conditions of HH~377. Figure 5 shows 
the flux-calibrated spectrum, extracted from a 4 $(\arcsec)^2$ in the E-W
oriented slit (see panel $b$ of Figure 4). The results obtained for the 
relative fluxes with and without reddening correction (for our three spectra)
are listed in Table 4. Column 1 gives the identification of the detected 
lines, column 2 lists the observed ($F$) and reddening-corrected ($F_0$) 
relative fluxes for the E-W slit, column 3 and 4 list the fluxes from
the N-S slit at low and intermediate resolutions, respectively.
The determination of the value of   $E(B-V)$ that we have used 
is discussed in the following section. In addition, the radial velocity
estimate from these spectra is $V_r =$-70$\pm 10$ km s$^{-1}$.

From the  spectrum with larger wavelength coverage (see Table 4 and Figure 4), 
we can see that [NI]5200 is greater than H$\beta$ and that the [SII]6717/31 
and [OI]6300/64 fluxes are stronger than H$\alpha$. No [OIII]5007 nor 
[NII]6548,83 emission is detected. These observed characteristics are typical
of low excitation Herbig-Haro objects.

Finally, we use the [SII] 6731/6717 line ratio to compute an electron density
$n_e=4100~cm^{-3}$ (assuming an electron temperature $T_e=10^4$ K).
As discussed in the following section, this result is particularly
interesting.

\section{Discussion: Cep E as a low excitation Herbig-Haro object}

As the limited wavelength coverage of our spectra does not allow us to
use Miller's (1968) method, we have determined the reddening by assuming 
that the intrinsic (dereddened) Balmer decrement has a H$\alpha$/H$\beta$=3, 
(i.e., a recombination cascade value). Using the measured 
H$\alpha$/H$\beta$=7.75 
ratio and the standard, $R_V= 3.1$~ ISM reddening curve of Mathis (1990), 
we obtain $E(B-V)=0.88\pm0.12$ (corresponding to an $A_V = 2.72\pm0.38$ 
optical extinction). This method is of course uncertain for the case of 
Cepheus E (and other HH objects) since it is expected that in the case 
of low velocity shocks the Balmer decrement could differ substantially from 
the recombination value. We have used the optical total-to-selective ratio 
$R_V= A(V)/E(B-V)= 3.1$ which is frequently used in this cases and appears 
to be the most appropriate for HH objects \markcite{boh91}.

In Table 5 we present line ratios observed for Cep E from the fluxes given 
in column 2 of Table 4.
The optical spectroscopic characteristics of
HH~377 (Cep E) identify it as a low excitation object. Indeed, we
find that the excitation of this object appears to be anomalously low,
compared to other low excitation HH objects. For example,
most of the observed line ratios obtained for HH~377 are consistent with
the ones observed in other low excitation objects,
except for the [SII](6717+6731)/H$\alpha$ ratio.
For HH~377, this ratio is a factor of $\sim 3$ higher than the corresponding
values for the  low excitation HH objects (HH~7, HH~11, HH34(jet), HH~47A,
 HH~111 D-J, HH~125 I and HH~128), this fact is clear in second panel top to
bottom, Figure 6.
 The [SII] 6731/6717 ratio is also the highest
one observed (with a value of $0.59\pm0.09$). This line ratio implies
an electron density $n_e \approx$4100~cm$^{-3}$, which is the largest
$n_e$ measured for any previously detected, low excitation HH object
(see Figure 6, bottom panel). The electron density is relatively high
and comparable to the densities measured in high excitation HH objects.
The low excitation in Cep E suggests, however, a low ionization fraction
and therefore a higher gas density than usual.

Following Raga et al. (1996), we compare the line ratios
of Cep E (HH~377) with those obtained for other HH objects.
We refer the line ratios to the [OI]6300/H$\alpha$ ratio, which is
different from Raga et al.(1996), who plot all of the line ratios versus the
[N I](5198+5200)/H$\beta$ line ratio. 
We have done this because the [N~I] lines have a very low critical density
($\sim$ 2900 cm$^{-3}$), so that collisional quenching will occur in Cep E 
(while not in the other low excitation HH objects).

In Figure 6 we plot the [NI](5198+5200)/H$\beta$, [SII](6717+6731)/H$\alpha$,
and [SII]6731/6717 line ratios (and also, the electron density $n_e$)
as a function of the [OI]6300/H$\alpha$ ratio, indicating the values
that correspond to high, intermediate and low excitation HH objects.
The line ratios for Cep E are indicated with an asterisk.
In this Figure we can see the anomalous line ratios, specially the
[SII](6717+6731)/H$\alpha$ ratio and electron density ($n_e$) 
obtained for Cep E, in contrast to other low excitation HH objects.

Let us now compare the line ratios of HH~377 (Cep E) with shock model
predictions.
The plane shock models of Hartigan et al. (1994) do show high
[SII](6717+6731)/H$\alpha$ ratios, in better agreement with the observations
of HH~377. From this work, we list the range of line ratios obtained for models
of shock velocities
in the v$_s=20$-30~km~s$^{-1}$ range (labeled J4.20-30 in Table 5),
and models with v$_s=30$-40~ km~s$^{-1}$ (labeled J4.30-40), in both cases
with a high pre-shock density n$_0=10^4$ cm$^{-3}$
and magnetic fields between 30 and 300 $\mu$G. 
We have also listed the line ratios  for models with v$_s=30-40$~km s$^{-1}$ 
and lower pre-shock density (n$_0=10^2$ cm$^{-3}$) and the same magnetic field
range than the models described above.
  
We can see that the J4.20-30 models are the most successful at reproducing
the line ratios observed in Cep E. In particular, these shock models
do reproduce the observed [SII](6717+6731)/H$\alpha$ and
[S~II]6731/6717 ratios.  Interestingly, these models also
predict an H$\alpha$/H$\beta~ \sim 6$ Balmer decrement. 

This result seems to favor shock models with a very low shock velocity
and high pre-shock density in order to reproduce the observed optical
line ratios. One has to keep in mind that J-type shocks in molecular
gas are restricted to velocities $\sim 30 - 50$~km s$^{-1}$ 
(Hollenbach \& McKee 1989) in order to not
dissociate H2, and so at first order the simple J-type
atomic/ionic plane parallel shock models are consistent with the rich
H2 spectra observed in Cep E.

We have estimated a new value for the extinction to Cep E using
the H$\alpha$/H$\beta~ \sim 6$ predicted by these models,  obtaining 
$E(B-V)\sim 0.24$ (corresponding to an $A_V \sim 0.77$ optical 
extinction).
However, it is important to note that for Cep E, only the brightest region of 
the southern lobe of the  H$_2$ outflow reveals optical emission. 
This fact suggests that a high extinction close the outflow region could
explain the observed
differences between the optical and IR morphologies.

Lefloch et al. (1996) determined values for the
optical extinction using observations of the continuum emission of
Cepheus E outflow at 1.25 mm.
They reported $A_V =3.2,3.4$ for the northern and southern H$_2$ lobes,
respectively (on the same bright infrared knots  which we analyze in this paper).
These values are similar to our first $A_V$ estimation from the
recombination cascade  Balmer decrement for HH377,
in the beginning of this section. The result implies a similar
extinction for both infrared lobes.
On the other hand, Noriega-Crespo et al. 1998 find a constant 
$v$ = 00 S(3)/0-0 S(5) ratio that also indicates the lack of a steep
extinction gradient between the north and south lobes and that the 
extinction is mostly important around IRAS 23011+6126. 
This is interesting because we know that the south lobe is visible at 
optical wavelengths (Noriega-Crespo 1997), while the north lobe is not.
It would be interesting to obtain extinction values for HH~377 through 
Miller's  method.
This would give a realistic value for the fraction of 
the extinction which arises in the vicinity of the object, and would
help to discriminate between and/or to constrain  the different shock models.

\subsection{The ionization fraction in HH~377}
 
In order to estimate the ionization fraction in the southern lobe
of Cep E (HH~377) we use the [OI]6300/H$\alpha$ ratio, which for
recombination H$\alpha$ is given by
\begin{equation}
{\rm{I([O~I]~6300}) \over I(H\alpha)}~=~\it{X}(\rm{O})~
{\rm{n(O~I)}/ n(O) \over n(H{^+})/ n(H)}~
{\it{q}_{ex} \rm{(6300)} \over \alpha (H \alpha)}\,,
\end{equation}
where ${X}(\rm{O})$ is the oxygen abundance, $q_{ex} \rm{(6300)}$ is the
excitation rate coefficient for [O~I]~6300, and $\alpha (\rm{H} \alpha)$
is the effective recombination rate coefficient for H$\alpha$.

For low excitation HH objects we expect electron temperatures lower
than 10$^4$ K (Bacciotti \& Eisl\"offel, 1999). If we assume that
T$_e ~ \sim$ 5000 K, we obtain $q_{ex} \rm{(6300)} \sim 4 \times 10^{-10}$
cm$^3$ s$^{-1}$,
and $\alpha (\rm{H} \alpha) \sim 2 \times 10^{-13}$ cm$^3$ s$^{-1}$.
For an oxygen abundance of ${X}(\rm{O}) = 8\times 10^{-4}$, and the
 ${\rm{I([O~I]~6300}) / I(H\alpha)}$ = 2.5 ratio  observed in HH~377,
from eq. (1) we obtain  ${\rm{[n(O~I)}/ n(O)]~/~[n(H{^+})/ n(H)]}~=$ 1.6;
which implies  that the gas is roughly 50$\%$ ionized.

On the other hand, we can also estimate the  ionization fraction,$x_e$,  
from the nitrogen lines using the ratio
\begin{equation}
{\rm{I([N~I]~5200})\over I([N~II]~6584)}~ \sim~ 0.13~ {n(N~I)\over n(N~II)}\,.
\end{equation}
Using the ${\rm{I([N~I]~5200})/ I([N~II]~6584)}~>~10$ limit deduced from the
spectrum of HH~377 (shown in Figure 5), we then obtain
${\rm{n(N~I)} / n(N~II)}~<~0.01$,
which implies that the ionization fraction is less than 1\%.
A similar value is obtained for the [SII]-weighted ionization fraction  using
the different line ratios in the diagrams shown in Hartigan et al.
(1994, Figures 3 to 5), for the models J4.20-30 (see above).

From the discrepancy between this low ionization fraction and the much higher
one deduced from eq. (1), we conclude that the H$\alpha$ line has to be 
collisionally  excited. We therefore conclude that the low reddening for 
HH~377 obtained through the shock models 
(in which the Balmer decrement is about 6), is probably correct.
If we consider the compression produced by the shock (which is approximately 
equal to the square of the  Mach number) and the ionization fraction 
deduced above we can estimate the pre-shock density. 
The shock velocity estimated for HH~377 (see section 4.0)
implicates a Mach number of $M^2 \sim 10$. 
Using an ionization fraction 
$x_e \sim 1\%$ and  considering that  $N_e \sim 10^4$ cm$^{-3}$,
we then obtain a pre-shock density ~$\sim$ 10$^5$ cm$^{-3}$. 
Note that this total particle density results very high at the distance
from the outflow source, and it is very unusual for HH objects with any
excitation level.

\subsection{The absolute fluxes in HH~377.}

 From the Hartigan et al. (1994)  shock models with 
$v=$ 20 km s$^{-1}$ and pre-shock density of 10$^4$ cm$^{-3}$, one obtains 
an H$\alpha$/H$\beta \sim 9$ and fluxes of 
H$\beta = 1.0 \times 10^{-5}$ and H$\alpha = 9.0 \times 10^{-5}$~
ergs cm$^{-2}$ s$^{-1}$, out of the front of the shock.
For a pre-shock density of 10$^5$ cm$^{-3}$ one would have fluxes larger by
about an order of magnitude.
This gives 
H$\beta = 1.6 \times 10^{-5}$ and H$\alpha = 1.5 \times 10^{-4}$~
ergs cm$^{-2}$ s$^{-1}$ steradian$^{-1}$, when applying the solid angle.
Using an area of 1.5" x 4" (see column 2 of Table 4), we obtain
that the predicted fluxes for H$\beta$ and H$\alpha$ are 
$2.3 \times 10^{-15}$ and $2.1 \times 10^{-14}$ergs cm$^{-2}$ s$^{-1}$,
respectively.
 These fluxes are about 6 times larger than the observed fluxes (see
Table 4).  However, the Balmer line fluxes of the Hartigan
et al. (1994) models  are extremely steep functions of shock speed, 
so a shock speed between 15 and 20 km s$^{-1}$ would give the correct 
total fluxes.

\vskip 0.9 truecm
Finally, we must ask how the optical emission and the molecular hydrogen 
emission are related in this dense, low excitation HH object.  Hartigan et 
al. (1996) discuss the possible morphologies.  One option is a bow
shock which has a J-shock nature at its tip (producing optical emission) 
with C-shocks or turbulent entrainment producing  the $\rm H_2$ emission 
from the bow shock wings.  Another is a C-shock in the ambient cloud 
accompanied by a J-shock at the Mach disk in the jet material.  
A third possibility is a J-shock with an MHD precursor which produces 
both optical and IR emission at the bow shock tip.  
Unfortunately, the high density of Cep E, the emitting 
regions extremely thin, so that it is difficult to use morphology to 
distinguish among the possibilities.
 
The various combinations of C- and J-shock models have  many free 
parameters, so that it is also difficult to use spectra to distinguish 
among them.  One interesting comparison, however, is the ratio between IR 
and optical luminosities. If the optical emission arises from the Mach disk 
and the $\rm H_2$ emission from the bow shock, and if the jet and ambient 
densities are not too drastically different, the luminosities should be 
comparable (e.g Hartigan 1989).  The C-shock luminosity emerges mostly in 
the near IR lines we observe, while according to the 
slow J-shock models, the luminosity is around 1000 times the H$\beta$
luminosity, with most of the energy emerging in Ly$\alpha$
(cf. the 15 and 20 $\rm km~s^{-1}$ models of Hartigan et al. 1994). Thus 
\begin{equation}
{\rm F_{IR} \sim 7\times10^{-12}  ~ergs~ cm^{-2}~ s^{-1} }\,,
\end{equation}
calculated from data in Table 2. And $\rm F(H\alpha) \sim 3\times 10^{-14} 
~ergs ~cm^{-2} ~s^{-1}$ for a smaller reddening than the 0.88 (Table 3), 
then we can estimated
\begin{equation}
{\rm F(Ly \alpha) \sim 1000 ~F(H\beta) \sim 150~ F(H\alpha) 
\sim 4.5\times 10^{-12} ~ergs ~cm^{-1}~ s^{-1} }\,.
\end{equation}
Thus it seems as though the J-shock is within an order of
magnitude of the C-shock luminosity, so the possibility that
the Mach disk makes the optical emission and the bow shock makes
the IR emission is reasonable.

\section{Conclusions}

We have carried out a spectroscopic and imaging study of the molecular 
hydrogen and optical atomic/ionized  emission in the  Cepheus E outflow. 
Our main results are:

\noindent
- For deriving the excitation state of molecular gas in Cep E we use line
ratios from our H-band and K-band spectra. 
We find  that T$_{exc}$= 2260$\pm$110 and 2340$\pm$100 K
for the northern and southern lobes, respectively, which are
consistent with the  T$_{exc}$ measured  in other HH objects. 
 The (1,0)/(2,1) S(1) ratios (8.50 and 9.40), and the Ortho/Para 
ratios  (values $\sim 3$)  in  both lobes are also consistent with the
values observed in collisionally excited objects.

\noindent
- Contrasting with the complex structure of the H$_{2}$ outflow,
the optical emission is a compact, well resolved  knot (HH~377),
that nearly coincides with the southern NIR lobe.
The [S~II] emission of HH~377 is clearly brighter and more compact
than  the H$\alpha$ emission and its angular size is
about 0.02 pc (at a distance of 0.75 kpc). The [S~II] and H$\alpha$
peak emission spatially coincide and appear offset a few arcseconds
upstream from the H$_2$ peak emission.

\noindent
- Our spectroscopic optical analysis reveals that  HH~377 has characteristics 
typical of low excitation Herbig-Haro objects.
This is confirmed when comparing the relative fluxes of HH~377 with 
those of  other HH objects using  line ratio diagrams. However, 
HH~377 presents anomalous [SII](6717+6731)/H$\alpha$ line ratio, larger 
than those obtained for objects classified as 
low excitation HH objects.  The electron density, $n_e$= 4100 cm$^{-3}$, 
determined for this object from [SII] lines would be the highest  density
measured in low excitation HH objects. This value is similar to the 
electronic densities measured in high excitation HH objects.

\noindent
-  We estimate an ionization fraction $x_e \sim 1\%$ for HH~377.
Together with the observed [OI]6300/H$\alpha$ ratio, this result
implies that the observed H$\alpha$ line has to be collisionally  
excited. This result supports the low reddening obtained for HH~377 
obtained through the shock models. Using this ionization fraction,
 a post-shock electron density $N_e \sim 10^4$ cm $^{-3}$ and a 
compression of $\sim 10$,  we obtain a pre-shock density ~$\sim$10$^5$ 
cm$^{-3}$~for HH~377. This exceptionally high pre-shock density
is very unusual for HH objects.

\noindent
- From the shock model predictions and the H$\beta$ and H$\alpha$
observed fluxes, we find that a shock speed between 15 and 20 
km s$^{-1}$ gives the correct  total fluxes for HH~377.
This velocity appears to be somewhat lower than the one deduced from 
the observed line ratios.

\noindent
- From a comparison between optical and infrared luminosities in HH~377
we find the possibility that the Mach disk produces the optical emission 
and the bow shock produces the IR emission.

\noindent
-  We have determined a visual extinction  $A_V = 2.72$ ($E(B-V)$=0.88)
assuming a recombination cascade H$\alpha$/H$\beta$=3 Balmer decrement.
If we use the H$\alpha$/H$\beta \sim 6$ decrement predicted by the 
preferred shock wave models we obtain an $A_V \sim 0.77$ ($E(B-V) \sim 0.24$)
extinction. Interestingly,  Lefloch et al. (1996) have obtained an 
$A_V = 3.4$ from the mm continuum of the southern lobe of the Cep E outflow, 
in qualitatively good agreement with the extinction obtained assuming 
H$\alpha$/H$\beta$=3 (see above).
This confusing situation involving the optical extinction towards 
HH~377 and the Balmer decrement predicted from low velocity shock models
 could be clarified with future observations of the blue and IR [SII] 
lines (or, alternatively, the IR [Fe II] lines) of this object, 
in order to have a model-independent determination of the extinction.
The overall extinction along the Cep E flow varies drastically. 
One could speculate different reasons (e.g. a large inclination with respect 
to the  plane of the sky, a dense molecular gas core surrounding the source 
with a rapidly decreasing density profile and/or a non-homogeneous ISM). 
We can not answer this question with our present data.

\noindent
- Comparing the {\it optical} line ratios observed in south lobe of Cep E 
to atomic/ionic plane-parallel shock models (J-type) presented in the 
literature (Hartigan et al. 1994), we find that low velocity (v$_s=20$-30) 
shocks are the closest to reproduce the observations. These models have a
high pre-shock density (n$_0=10^4$ cm$^{-3}$), and 
magnetic fields between 30 and 300 $\mu$G. These conditions (high density,
low ionization fraction and strong magnetic field) are quite appropriate
for the development of {\t molecular} C-type shock as well. 
Previous comparisons of the H$_2$ near-ir spectra with molecular shocks 
(Ladd \& Hodapp, 1997),
indeed indicated a preference for C-type shocks with $\sim 35$ \kms.
Preliminary results from ISO Long Wavelength Spectrometer (50 - 200\mum)
support also this view, given Cep E rich H$_2$O spectra (Noriega-Crespo 2000),
a characteristic signature for C-type shocks (Kaufman \& 
Neufeld 1996ab; Noriega-Crespo et al. 2000)

\noindent

 - Finally, our analysis seems to confirm that Cep E corresponds to 
an outflow in its earliest developing phases. Its short dynamical age of few 
10$^3$ yrs, the high gas density estimated (at least 10$^5$ \cc) and the 
inhomogeneous nature of its extinction, suggest that the outflow is breaking 
through its placental molecular core.

\acknowledgments
This work was supported by DGAPA grant IN109297 and CONACYT grant
26833-E and 27546-E.
S.A. \& S.C. would like to thank C\'esar Brise\~no for obtaining some 
of the optical images.
A.N.-C. \& P.M.G thank Nichole King for her help on the observations. 
The FSPEC IR spectra were obtained with the help of Marcia Rieke \&
George Rieke at the MMTO,  A.N.-C. \& P.M.G are greatful to both.
K.H.B. has been supported by NSF grant AST 9729096. Last but not least,
we thank the referee David Devine for his comments and careful reading
of our manuscript.

\clearpage
\plotone{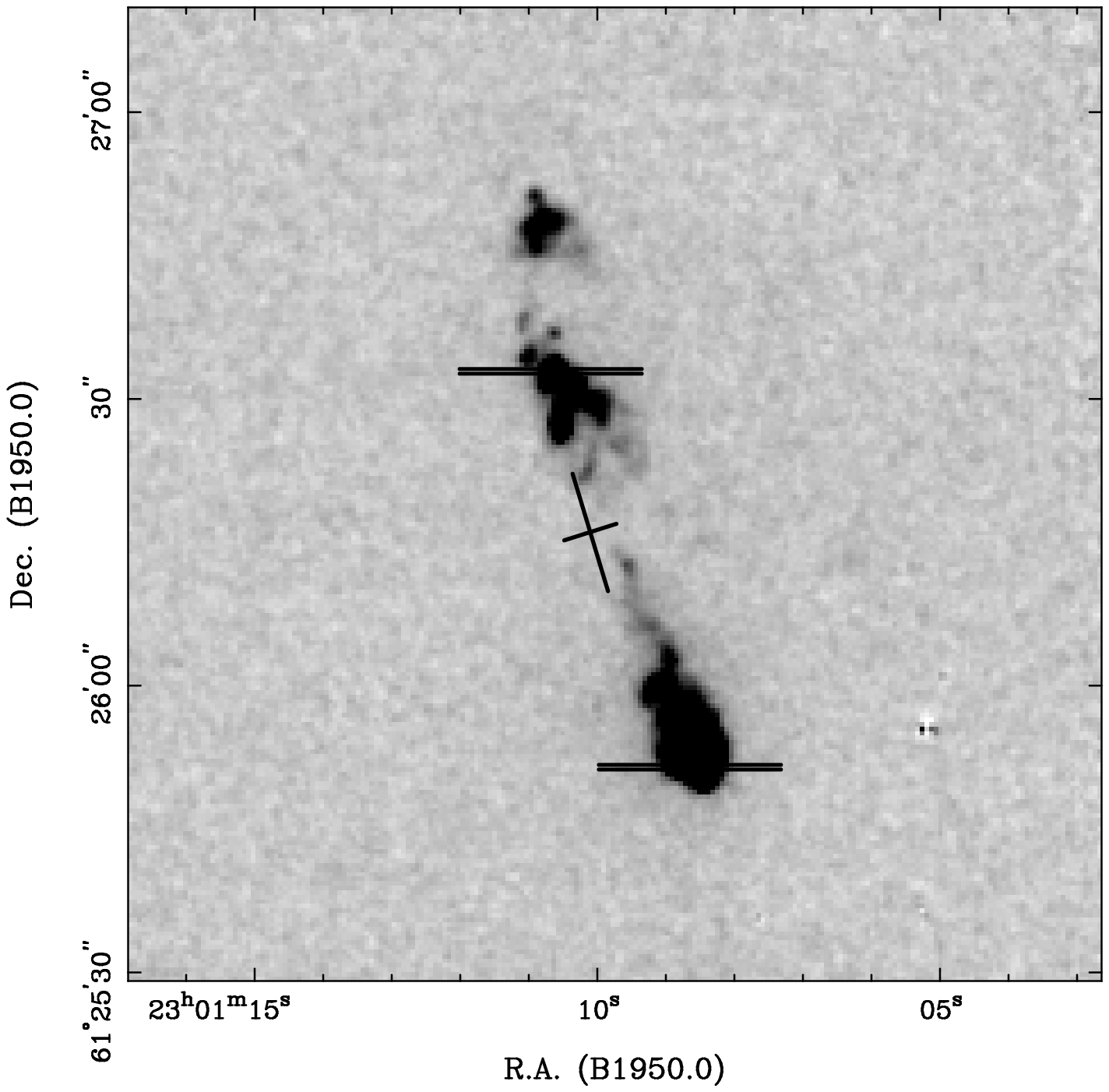}
\figcaption[cepren.ps]{Grey-scale subsection of the H$_{2}$ at 2.12 $\mu$m 
 continuum subtracted image around Cepheus E outflow region.The cross 
indicates the position of IRAS23011+6126 and it has a size comparable
to that of IRAS uncertainty. The schematic slit positions for the 
infrared spectra discussed in the text are shown.\label{fig_ir}}

\plotone{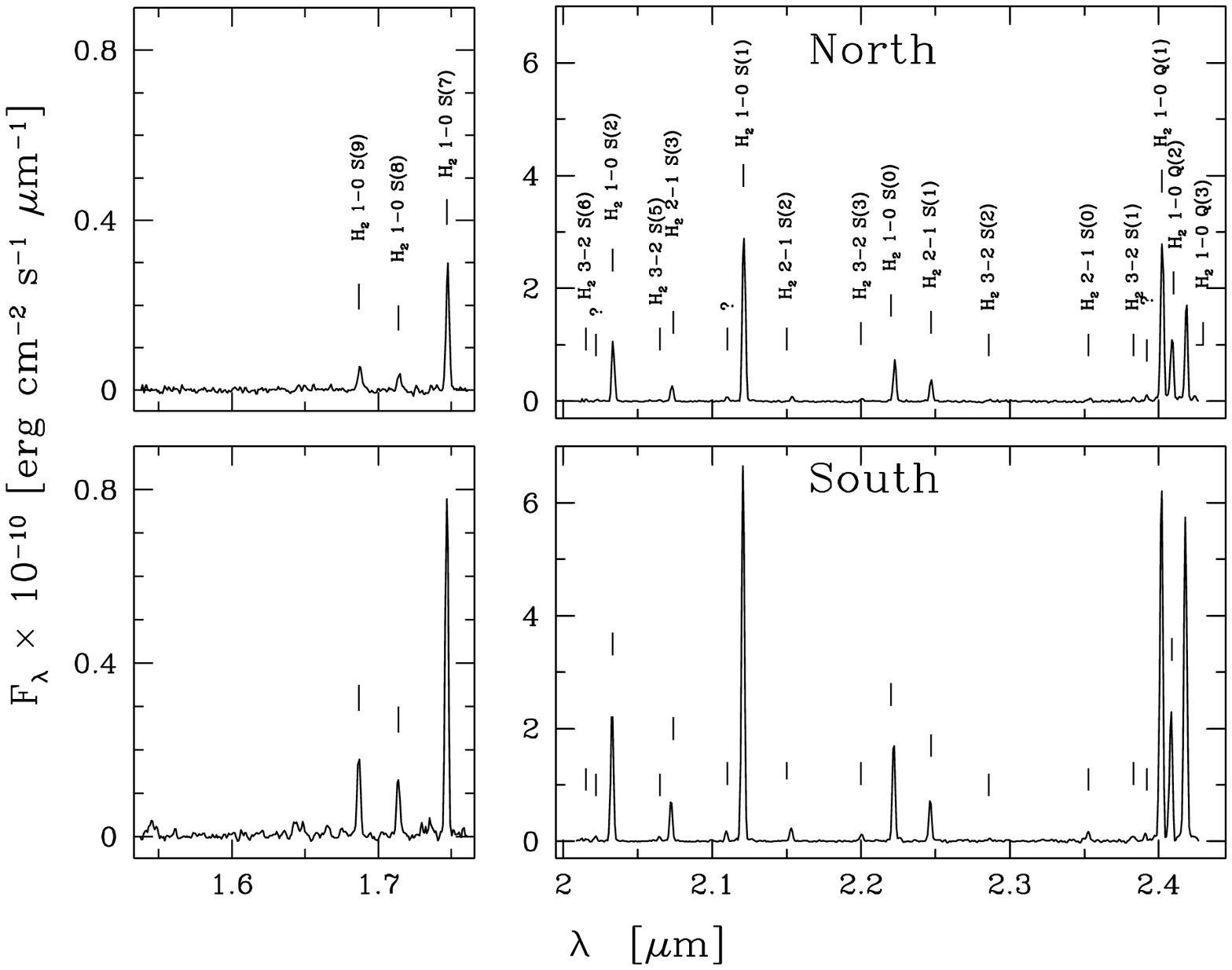}
\figcaption[cepespecmu_1.ps]{H and K band calibrated spectra of brightest 
knots of Cepheus E
outflow. In ({\it top panel}) for 
northern lobe and ({\it bottom panel}) for the southern lobe.
The wavelengths of the expected   H$_2$ lines are marked.
Note that Br $\gamma$ line emission
is not detected.
\label{fig_irspec}}

\plotone{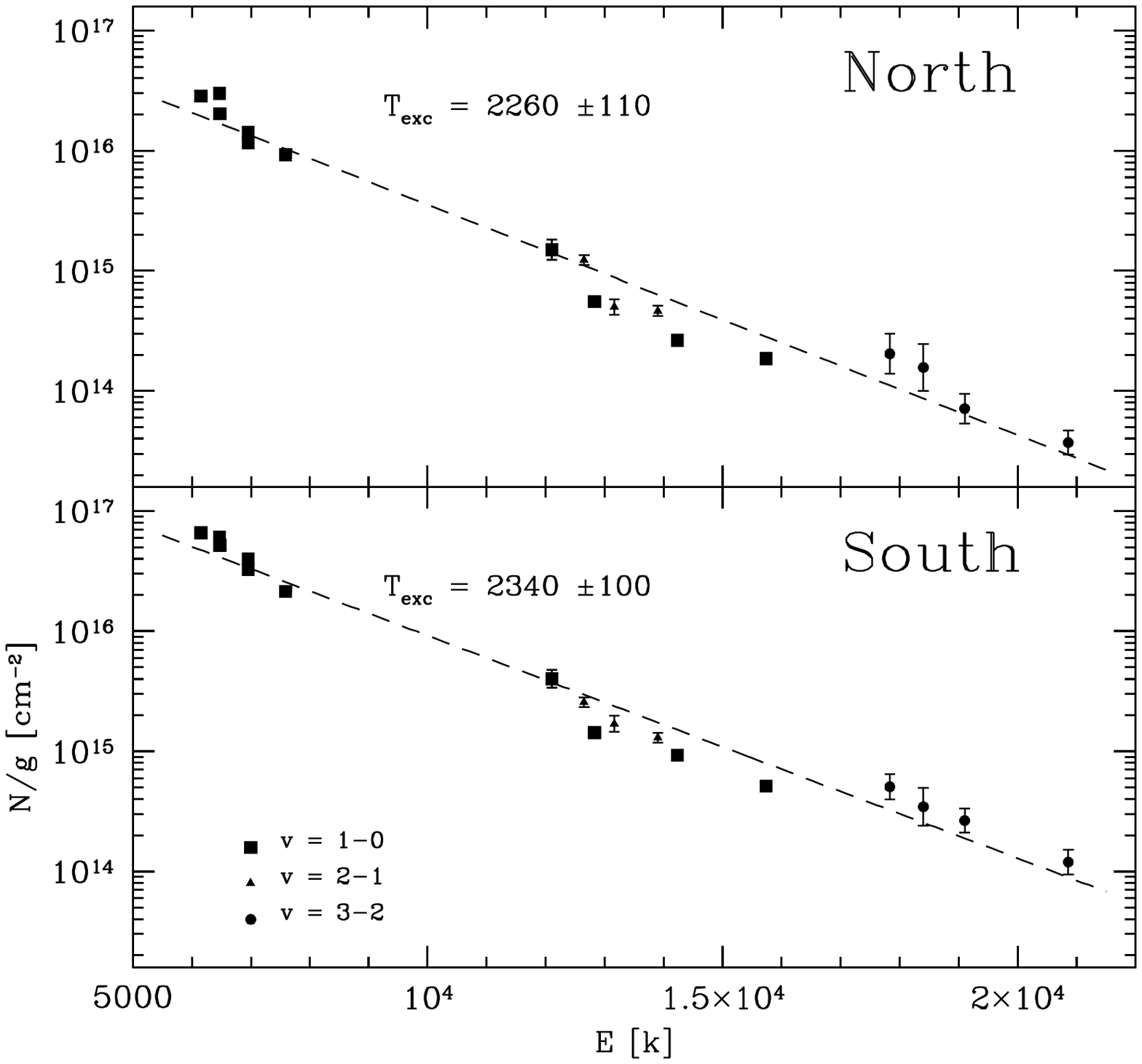}
\figcaption[cepeTexc.ps]{Excitation diagrams for the position on the 
brightest 
knots in Cepheus E lobes (shown in Figure 1). Different symbols have been 
employed for each  vibrational level and are indicated in the bottom panel.
The dashed lines are the least-square fit to all plotted transitions
for the Northern lobe ({\it {upper panel}}) and the Southern  lobe
({\it {bottom panel}}). The excitation temperature calculated for 
 the fits is also indicated.
\label{fig_texc}}

\plotone{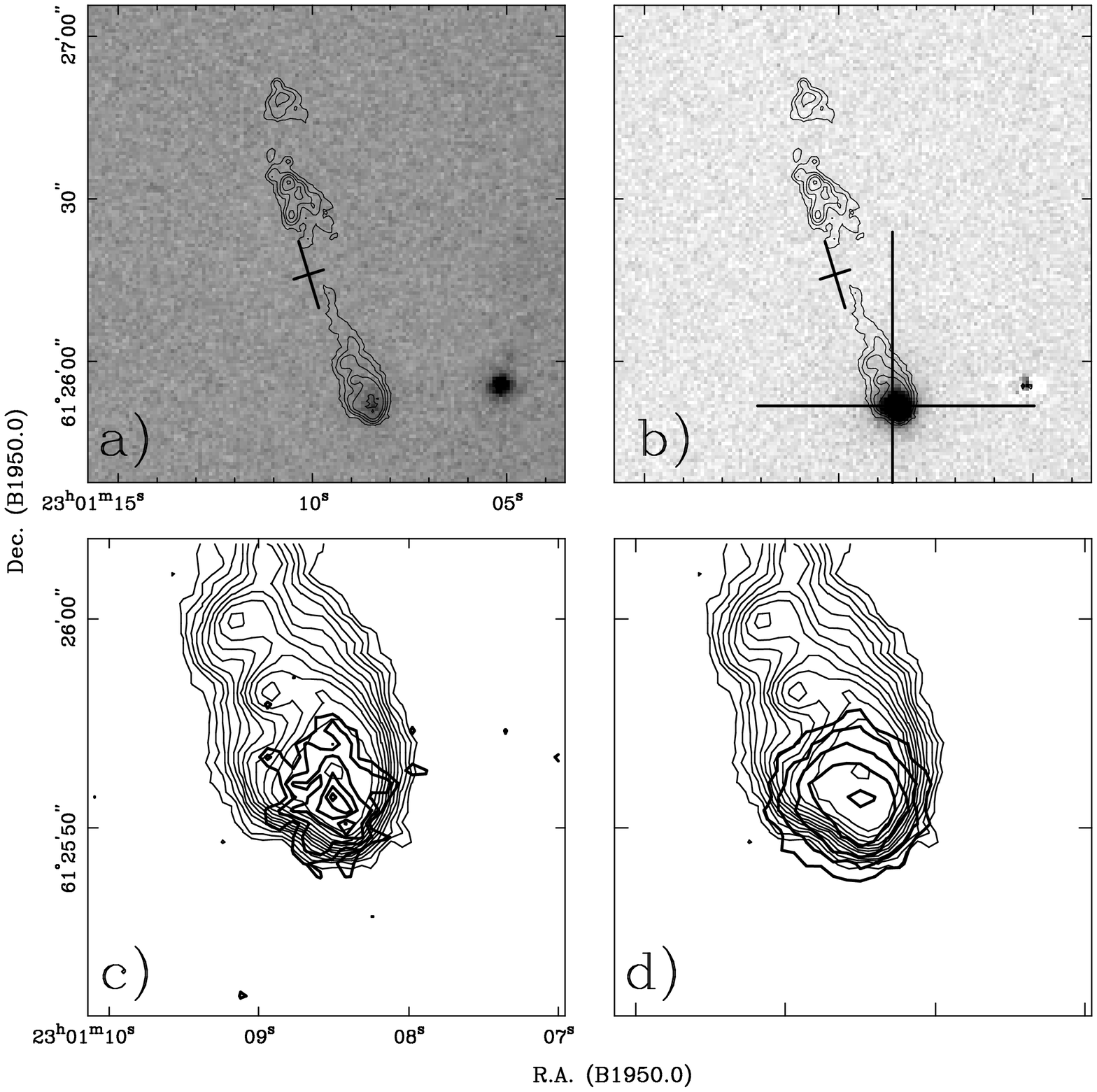}
\figcaption[fig4consub_1.ps]{
{\bf a)} Contour map for the H$_2$ (1,0) S(1) continuum 
subtracted image (shown in Figure \ref{fig_ir}), 
overlaid on a grey-scale subsection 
of the  H$\alpha$ continuum subtracted image of the region around 
Cepheus E.  The contour interval is a logarithmic factor of 0.36. 
The cross indicates the IRAS23011+6126 position.
{\bf b)} The same H$_2$ (1,0) S(1) contour map (panel a) now
 overlaid on the grey-scale [S~II] \lam\lam6717/31 continuum 
subtracted image.
This map shows the same image size than panel a). The  right lines 
are the schematic slit positions for the optical spectra discussed
in this work.
{\bf c)}An overlay of the contour plots of the emission from a 
close-up  of Cepheus E southern lobe  showed in panel a (continuum 
subtracted). The dark 
contours represent the  H$\alpha$ emission; the light contours
represent the  H$_2$ (1,0) S(1) emission. The contour interval
increases linearly by a factor of the 1.1 for the H$\alpha$ map,
 and by a logarithmic factor of 0.13 for the  infrared map.
 Note the  spatial difference  between 
the emission peak in   H$\alpha$  and  the corresponding  peak 
in H$_2$ (1,0) S(1).
{\bf d)} An overlay of the contour plot of the images  showed in 
panel b. The dark  
contours represent the [S~II] \lam\lam6717/31 emission; the light 
contours represent the  H$_2$ (1,0) S(1) emission. In this case 
the contour interval for the [S~II]~is a logarithmic factor of 0.17. 
This figure shows a clear offset between the [S~II] and the H$_2$
peak emissions.
\label{fig_op4}}

\plotone{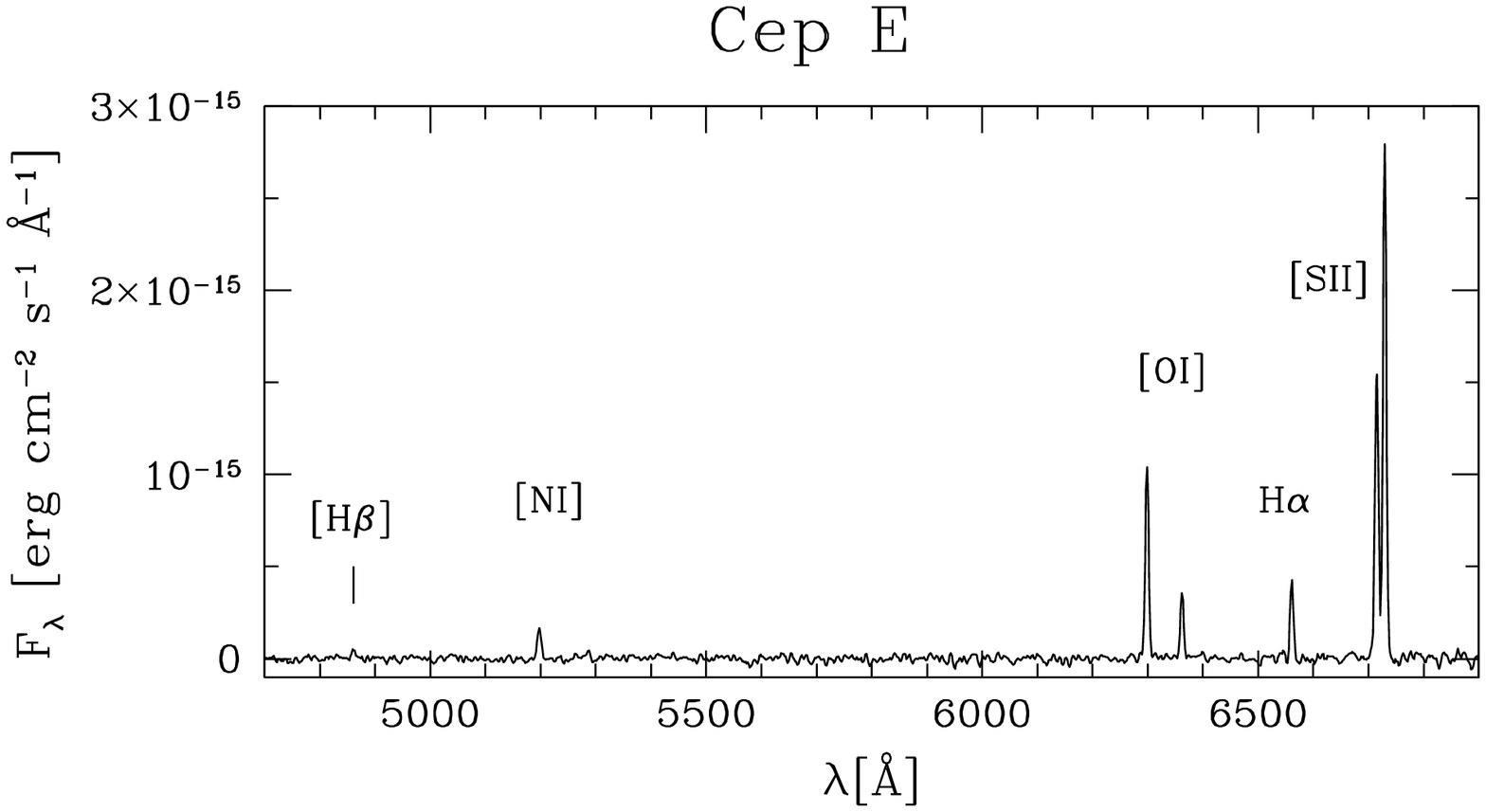}
\figcaption[cepeAPO600all.ps]{Low resolution spectrum of the
Southern lobe in  Cepheus E (named HH~377). 
The slit was E-W oriented as shown in Figure \ref{fig_op4} 
(panel b). The measured fluxes are listed in column (2) 
on Table~ 4.
\label{fig_opspec}}

\plotone{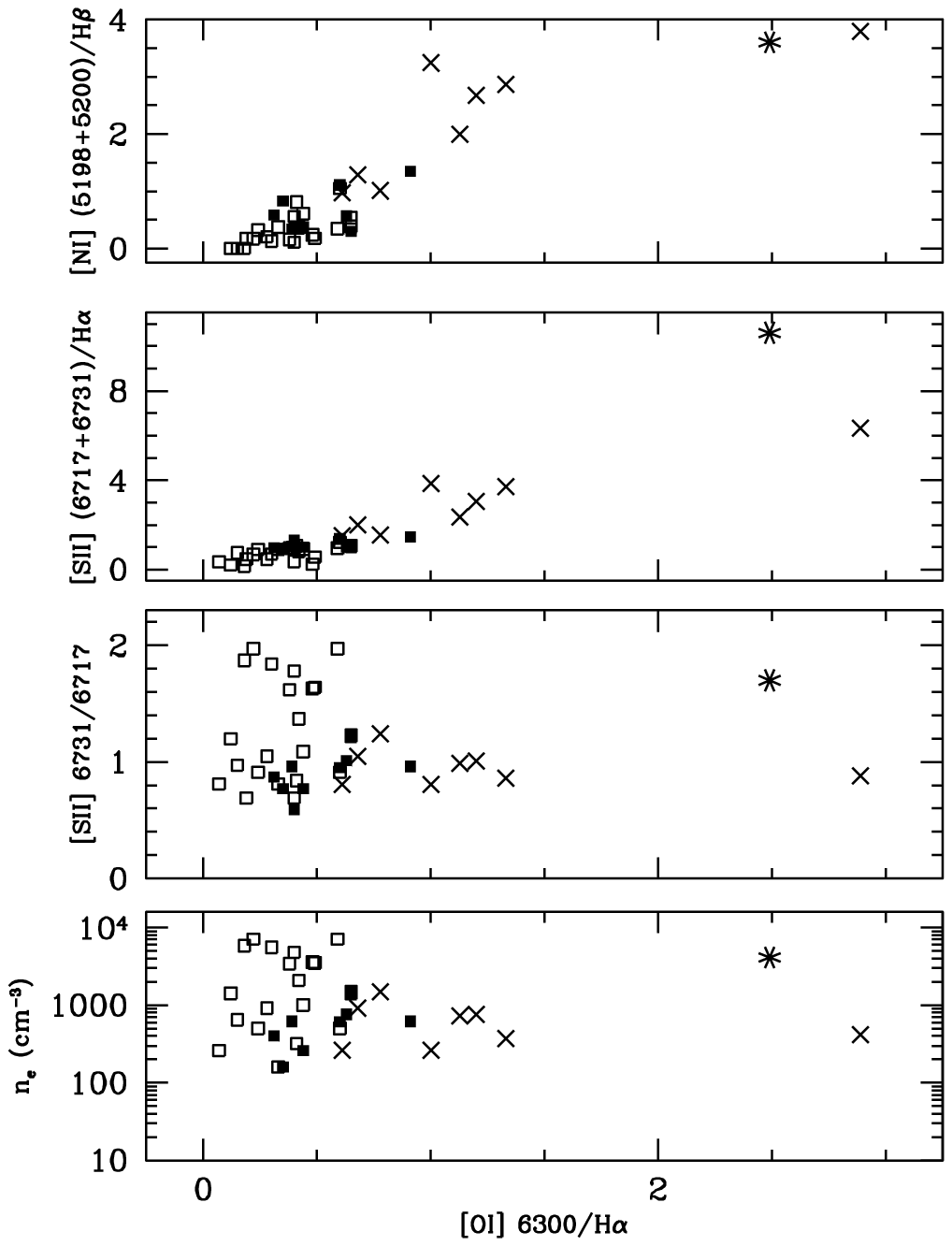}
\figcaption[hh_a.ps]{[NI](5198+5200)/H$\beta$, 
[SII](6717+6731)/H$\alpha$ and 
[SII]6731/6717 line ratios and electron density ($n_e$) as a function 
of [OI]6300/H$\alpha$ ratio. The objects
with high excitation spectra are shown with open squares, the intermediate 
excitation spectra with solid squares, the low excitation objects with crosses,
and Cepheus E with an asterisk.
\label{fig_hh}}

\end{document}